\documentclass[
twocolumn,
aps,
prb,
reprint,
superscriptaddress,
amsmath,
amssymb,
longbibliography
]{revtex4-2}
\usepackage[hidelinks]{hyperref} % pdfLatex
\usepackage[pdftex]{graphicx, color}
\hypersetup{% hyperrefオプションリスト
 setpagesize=false,
 bookmarksnumbered=true,%
 bookmarksopen=true,%
 colorlinks=true,%
 linkcolor=blue,
 citecolor=blue,
 urlcolor=blue
}
\usepackage{braket}
\usepackage{times,multirow,amsfonts,bm,xspace,pifont,mathtools}
\usepackage[normalem]{ulem}
\usepackage{soul}
\begin{document}
\newcommand{\rr}{{\bm r}}
\newcommand{\q}{{\bm q}}
\renewcommand{\k}{{\bm k}}
\newcommand*\YY[1]{\textcolor{red}{#1}}
\newcommand*\JI[1]{\textcolor{magenta}{#1}}
\newcommand{\JIS}[1]{\textcolor{cyan}{\sout{#1}}}
\newcommand{\YYS}[1]{\textcolor{blue}{\sout{#1}}}

% Title of paper
\title{Field-angle dependence of magnetoresistance in \texorpdfstring{UTe$_2$}{UTe2}} %\texorpdfstring{\\}{} title .. second row}

\author{Jun Ishizuka}
% \email[]{ishizuka@eng.niigata-u.ac.jp}
\affiliation{Faculty of Engineering, Niigata University, Ikarashi, Niigata 950-2181, Japan}

\author{Youichi Yanase}
% \email[]{yanase@scphys.kyoto-u.ac.jp}
\affiliation{Department of Physics, Graduate School of Science, Kyoto University, Kyoto 606-8502, Japan}
\date{\today}

\begin{abstract}
% insert abstract here
We theoretically study angle-resolved magnetoresistance under rotated magnetic field in the normal state of a spin-triplet superconductor UTe$_2$.
The Wannier model derived from a GGA+$U$ calculation shows quasi-two-dimensional Fermi surfaces with warping in the $k_z$ direction, consistent with quantum oscillation measurements in the high magnetic field regime.
Solving the semiclassical Boltzmann equation, we show that the Fermi surface geometry gives rise to oscillations in the magnetoresistance when the field is tilted from the $c$ axis toward the $a$ or $b$ axis. 
By assuming a band-dependent relaxation time, the calculated angle-resolved magnetoresistance is in good agreement with the recent transport experiment.
This is direct evidence for the warped Fermi surface revealed by ordinary intraband transport. It suggests that the hole band with long relaxation time dominates electron transport.
The field angle dependence of the Hall resistivity is calculated for further experimental verification.

\end{abstract}

\maketitle

\section{Introduction}
Superconductivity in the heavy fermion compound UTe$_2$ \cite{Ran2019Nearly,Aoki2019Unconventional,Aoki2022UTe2review} has attracted intense attention due to their exotic properties \cite{Knebel2019Field-Reentrant,Ran2019Extreme,Tokunaga2019125Te-NMR,Knafo2019Magnetic-Field,A.Miyake2019Metamagnetic,Thomas_UTe2_2020,Bae2021Anomalous,Sundar2019Coexistence,Duan2020Incommensurate,Knafo2021Low,Raymond2021Feedback,Ishihara2023Chiral,Tokunaga2023Longitudinal,Matsumura2025Intrinsic}, including the multicomponent spin-triplet superconductivity \cite{Braithwaite_UTe2_2019, Ran_UTe2_pressure, Lin_UTe2_2020, Knebel_UTe2_2020, Aoki_UTe2_2020, Kinjo2023Change,Tokiwa2022Stabilization} and the potential of topological superconductivity \cite{Ishizuka2019Insulator,Jiao2020Chiral,Wu2021Topological,Yu2022Majorana,Shaffer2022Chiral,Tei2023Possible}.
Topological superconductors accompanied by Majorana fermion excitations hold considerable promise for quantum computing applications \cite{Qi_review,Sato_review2016,Sato_review2017}.
Evidence supporting spin-triplet pairing in UTe$_2$ includes extraordinarily high upper critical fields \cite{Ran2019Nearly}, reentrant superconductivity near metamagnetic transitions \cite{Knebel2019Field-Reentrant,Ran2019Extreme}, and slight decreases in spin susceptibility below the superconducting transition temperature $T_{\rm c}$ \cite{Nakamine2019Superconducting,Nakamine2021Anisotropic,Nakamine2021Inhomogeneous,Fujibayashi2022Superconducting,Matsumura2023Large,Kitagawa2024Clear}. 
However, despite substantial efforts by theories and experiments \cite{Hayes2021Multicomponent,Wei2022Interplay,Ajeesh2023Fate,Wang2025Imaging,Sharma2025Observation,Shick2019Degeneracy,Xu2019QuasiTwoDimensional,Shishidou2021Topological,Hiranuma2021Paramagnetic,Ishizuka2021Periodic,Moriya2022Intrinsic,Kreisel2022Spin,Kanasugi2022Anapole,Chazono2023Piezoelectric,Kitamura2023Quantum,Rosuel2023Field,Theuss2024Single-component,Hazra2024Pair,Hakuno2024Magnetism,Zhang2025Dimensionality,Hakuno2026Superconductivity,Shimizu2025Magnetic,Hayes2025Field-angle-resolved,Totsuka2026Nodal}, the electronic structure, pairing mechanisms, and symmetry of superconductivity remain unsolved.

Intensive studies have been dedicated to the identification of the Fermi surface (FS) \cite{Fujimori_UTe2,Miao_UTe2,Niu_UTe2,Aoki2022First,Aoki2023de,Broyles2023Revealing,Eaton2024Quasi-2D}, which is crucial to a comprehensive understanding of exotic phenomena in UTe$_2$. 
First-principles calculations have predicted a wide range of underlying band structures \cite{Aoki2019Unconventional}.
Although the band structures far below the Fermi level are consistent between naive band calculations and angle-resolved photoemission spectroscopy (ARPES) measurements \cite{Fujimori_UTe2}, first-principles calculations predict Kondo-insulating-like bands, which contradict the metallic behavior of UTe$_2$.
Later, the GGA+$U$ calculation and the density functional theory combined with dynamical mean field theory (DFT+DMFT) find quasi-two-dimensional (2D) hole and electron FSs in agreement with ARPES \cite{Ishizuka2019Insulator,Xu2019QuasiTwoDimensional,Miao_UTe2}.
High-quality single crystals with $T_{\rm c}\sim2.1$ K were recently synthesized \cite{Sakai2022Single,Aoki2024Molten}, and thereby
the de Haas–van Alphen (dHvA) effect has been observed. The angular dependence of the dHvA frequencies revealed the warping of cylindrical 2D FSs in good agreement with the GGA+$U$ calculation \cite{Aoki2022First}.
The existence of three-dimensional (3D) FS was reported by Shubnikov-de Haas (SdH) oscillations \cite{Broyles2023Revealing} and implied by $c$-axis transport \cite{Eo2022c-axis}. 
Several theoretical studies also predicted a small 3D FS centered at the $\Gamma$-point \cite{Fujimori_UTe2,Choi2024Correlated,Kang2025Coexistence}.
However, such a 3D FS was not detected in recent dHvA and SdH measurements \cite{Aoki2024High}.

Quantum interference oscillations (QIOs) in magnetoresistance and magnetoconductance are a useful tool to investigate the bulk electronic structure \cite{Weinberger2024Quantum,Husstedt2025Slow}.
Unlike quantum oscillations from the dHvA and SdH effects, the QIOs are attributed to the semiclassical orbital motions of quasiparticles connecting separate FS sections. Therefore, QIOs and quantum oscillations can be complementary probes of the FS topology.
In UTe$_2$, the appearance of QIOs for magnetic fields along the $a$ and $c$ axes has been reported \cite{Weinberger2024Quantum,Husstedt2025Slow}.

Angle-resolved magnetoresistance is also informative for detecting the FS topology \cite{Yamaji1989Angle,Lewin2018Angle,Zhang2019Magnetoresistance}. 
A recent experiment of UTe$_2$ shows a monotonic increase of the $c$-axis resistivity $\rho_{zz}$ when the magnetic field is rotated from the $c$ axis to the $b$ axis, while it shows a nonmonotonic behavior with dips when the field is rotated toward the $a$ axis \cite{Kimata2026Electron-Hole}.
These behaviors of the angle-dependent magnetoresistance may be attributed to the warping of FSs. However, an explicit calculation has been lacking.

Inspired by the experimental transport studies, we investigate the angle-resolved magnetoresistance with a microscopic Wannier model derived from the first-principles band structure calculations with the warped 2D FSs.
By introducing the band dependence of the relaxation time, which is compatible with the antiferromagnetic fluctuation with the wave vector $\bm q \sim (0,\pi,0)$ observed in the neutron scattering experiments \cite{Duan2020Incommensurate,Knafo2021Low}, it is shown that the magnetic field angular dependence of resistivity reported in Ref.~\onlinecite{Kimata2026Electron-Hole} is explained by the majority contribution from the warped hole FS with small Fermi velocity.
This is direct evidence for the warped FSs revealed by ordinary intraband transport.
%, different from QIOs, dHvA, and SdH frequencies due to the quantum interference or Landau quantization.

\section{Method}
To calculate angle-resolved magnetoresistance, the electronic structure is described by density functional theory using the Wien2k code \cite{blaha_2}.
To include the correlation effects on uranium atoms, we adopt a GGA+$U$ method implemented in Wien2k\xspace.
We use $U = 2.0$ eV in the GGA+$U$ electronic structure calculation. This value is chosen because the resulting Fermi surfaces are consistent with those inferred from the quantum oscillation measurements \cite{Aoki2022First}. We note that the Fermi-surface topology depends on the choice of $U$~\cite{Ishizuka2019Insulator}, and therefore the calculated field-angle dependence of the resistivity can also be affected by $U$. The changes in the Fermi-surface geometry are expected to modify or smear the oscillatory behavior in the angle-dependent magnetoresistance.

We construct a Wannier model using Wannier90 \cite{Pizzi2020} through the wien2wannier interface \cite{Kunes2010} for $U=2.0$ eV.
It is beneficial to reproduce the electronic states near the Fermi energy with a small number of Wannier orbitals to reduce a calculation time.
Thus, we set projectors from the Bloch states to the Wannier functions as 5$f_{z^3}$ and 6$d_{3z^2-r^2}$ orbitals on the two U sites and 5$p_{y}$ orbitals on the two Te(2) sites [Fig.~\ref{fig:model}(a)] with spin degree of freedom, resulting in a 12-band model.
We perform a 100-times Wannierization with $12 \times 12 \times 12$ $\bm k$-points.
For comparison, we also constructed a 72-band Wannier model.

The conductivity/resistivity tensor is calculated using the Boltzmann equation with the semiclassical and relaxation time approximation as implemented in the WannierTools package \cite{Wu2018WannierTools,Zhang2019Magnetoresistance}.
Here, we briefly look at the formulas \cite{Ashcroft1976}.
The conductivity tensor with the magnetic field $\bm B$ is given by
\begin{align}
    \sigma_{i j}^{n}(\bm{B})=\frac{e^2}{4 \pi^3} \int d \bm{k} \tau_n v_i^n(\bm{k}) \overline{v}_j^n(\bm{k})\left(-\frac{\partial f}{\partial \varepsilon}\right)_{\varepsilon=\varepsilon_n(\bm{k})},
    \label{eq:Boltzmaneq}
\end{align}
% Note: quantum interference is not considered
where $\tau_n$ is the relaxation time at band $n$, $f$ is the Fermi-Dirac distribution function, $v_i^n(\bm{k})$ is the velocity for $i=x, y, z$ direction, defined by
$\bm{v}^n(\bm{k})=\nabla_{\bm{k}} \varepsilon_n(\bm{k})/\hbar$.
% \begin{align}
%     \bm{v}^n(\bm{k})=\frac{1}{\hbar} \nabla_{\bm{k}} \varepsilon_n(\bm{k}),
% \end{align}
In the above formula, ${\overline v}_i^n(\bm{k})$ is the weighted average of velocity over the past history of the electron orbital motion
\begin{align}
    \overline{\bm{v}}^n(\bm{k})=\int_{-\infty}^0 \frac{d t}{\tau_n} e^{\frac{t}{\tau_n}} \bm{v}^n(\bm{k}_n(t)),
\end{align}
with the time evolution of the wave vector $\bm k_n(t)$,
\begin{align}
    \frac{d \bm{k}_n(t)}{d t}=-\frac{e}{\hbar} \bm{v}^n(\bm{k}(t)) \times \bm{B}.
    \label{eq:k}
\end{align}
The Runge–Kutta method is adopted to solve Eq.~(\ref{eq:k}).
The total conductivity tensor is obtained by summing the contributions of the bands, ${\hat \sigma}(\bm B) = \sum_{n}{\hat \sigma^{n}(\bm B)}$.
The resistivity tensor is given by the inverse of the conductivity tensor ${\hat \rho}(\bm B) = [\sum_{n}{\hat \sigma^{n}(\bm B)}]^{-1}$.
In the semiclassical formula Eq.~(\ref{eq:Boltzmaneq}), the temperature dependence is introduced through the Fermi distribution function. The relaxation time $\tau_n$ is treated as a phenomenological band-dependent parameter, and its temperature dependence is not explicitly included. In principle, $\tau_n$ can depend on temperature through microscopic scattering processes, such as impurity, phonon, or magnetic-fluctuation scattering.

Note that in the semiclassical approximation, interband effects such as QIOs due to magnetic breakdown are neglected. We also assume that the FSs are not affected by the Zeeman field.
These semiclassical treatments are sufficient for the purposes of the present work, but these effects can influence the quantitative magnitude and fine structure of the field-angle dependence, especially in the higher magnetic field region.
% We set the relaxation time for hole and electron FSs as $\tau_{\rm h}=1$ ps and $\tau_{\rm e}=1$ and $0.1$ ps, and 
In the numerical calculations, $27 \times 27 \times 27$ $\bm k$-points are used.
% In Supplemental Material \cite{suppl}, we show the results for $\tau_{\rm h}=1$ ps and $\tau_{\rm e}=0.25$ ps.

\section{Result}
\label{sec:result}
\subsection{Electronic structure}
\begin{figure}[tbp]
\includegraphics[width=1.0\linewidth]{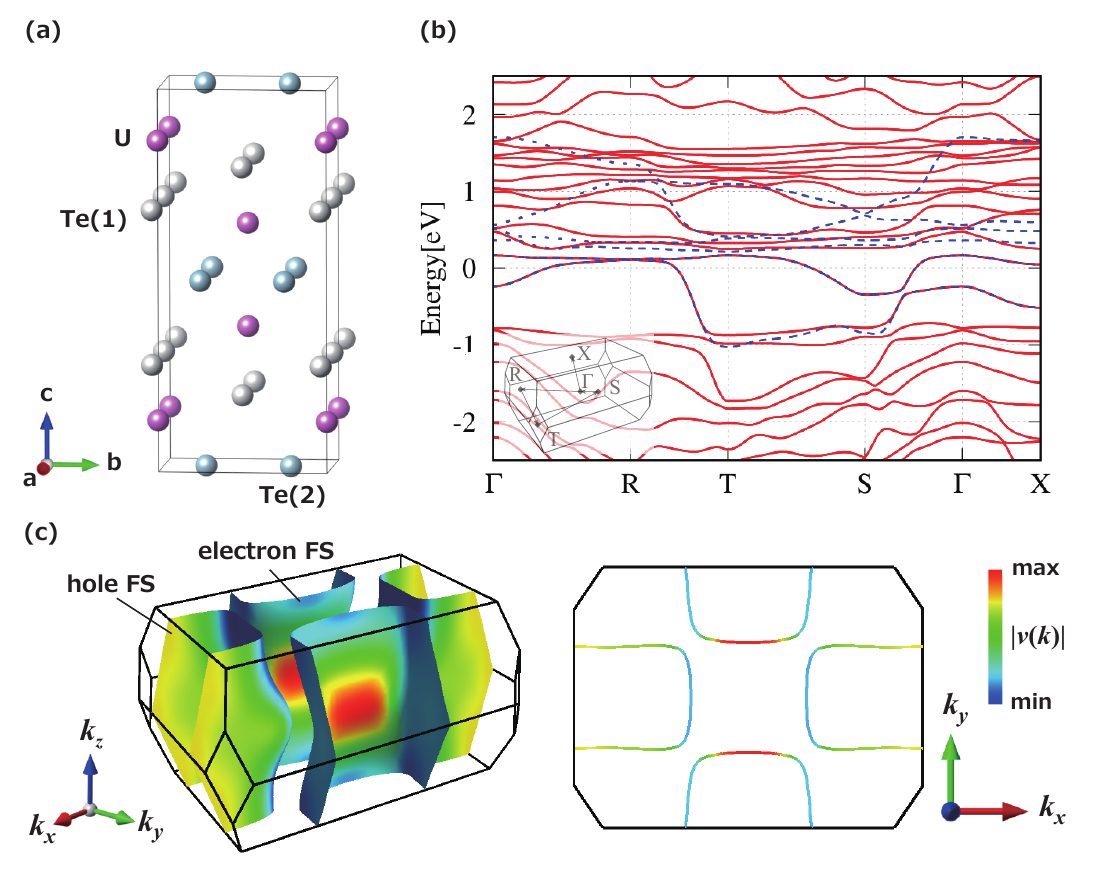}
\centering
\caption{(a) Crystal structure of UTe$_2$. (b) Band structure of the GGA+$U$ calculation for $U=2.0$ eV (solid lines) and the 12-band Wannier model (dashed lines) along the high symmetry lines. (c) Left: Fermi surfaces of the 12-band Wannier model. Right: 2D cut on the $k_z =0$ plane. The magnitude of Fermi velocity is indicated by color.
\label{fig:model}}
\end{figure}

Figure~\ref{fig:model}(b) shows the band structure obtained in the GGA+$U$ calculation for $U=2.0$ eV and that of the 12-band Wannier model. %along the high symmetry lines. 
We see that the Wannier model reproduces the low-energy states near the Fermi energy.
The FSs of the Wannier model are illustrated in Fig.~\ref{fig:model}(c), which shows two rectangular FSs.
The quasi-2D electron sheet shows warping near the $k_y=\pm\pi/2$ plane around the $S$-point. A large Fermi velocity is obtained in the warped region, while the Fermi velocity is small in the flat plane around $k_x \sim \pm \pi/2$.
These features are attributed to the one-dimensional (1D) dispersion of the $d_{3z^2-r^2}$ orbitals along the $k_x$ direction and the 1D dispersion of the $p_{y}$ orbitals along the $k_y$ direction.
As a consequence of finite hybridization between these bands and U 5$f$-electrons, the electron FS shows a shape similar to a warped rectangular tube with different velocity between the $k_x$ and $k_y$ axes.
The hole FS sheet can be understood in a similar way. 
%It should be noted that the situation is opposite for the hole sheet: 
However, unlike the electron FS sheet, the warping appears around the $k_x \sim \pm\pi/2$ plane, and the Fermi velocity is small there.

The band structure of the 12-band Wannier model is in good agreement with the 72-band Wannier model not only for the shape of the FSs but also for the Fermi velocity on the FSs (See Appendix \ref{ap:wannierfit} for details of Wannier fit depenence). %$is in good agreement with the realistic 72-band Wannier model.
Therefore, the 12-band Wannier model is a reasonable basis for the theoretical study of intraband transport.
The FSs are consistent with the quantum oscillation experiments \cite{Aoki2022First,Eaton2024Quasi-2D}, and ARPES has also observed quasi-2D FSs \cite{Miao_UTe2}.

\subsection{Magnetoresistance with band-independent relaxation time}
\label{sec:result_2}

\begin{figure}[tbp]
\includegraphics[width=1.0\linewidth]{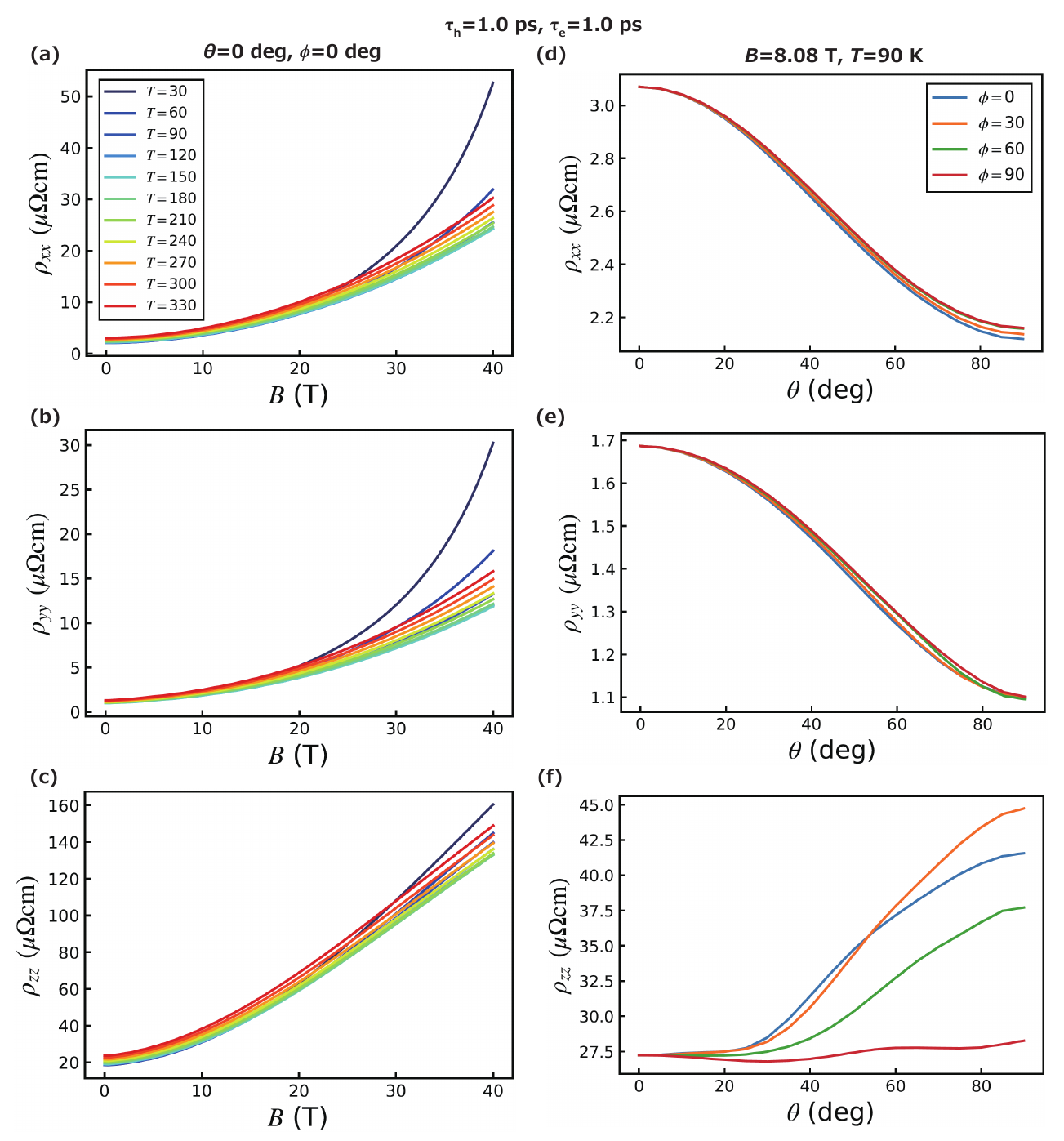}
\centering
\caption{(a)-(c) Magnetic field $B$ dependence of diagonal resistivity, namely, the magnetoresistance, for the field angle $\theta=\phi=0^\circ$. The temperature is varied from $30$ K to $330$ K. (d)-(f) Field angle $\theta$ dependence of resistivity at various angles $\phi$ for $B=8.08$ T and $T=90$ K. 
A band-independent relaxation time $\tau_{\rm h}=\tau_{\rm e}=1.0$ ps is assumed.
\label{fig:rho-diag_tau1.0}}
\end{figure}

The resistivity tensor as a function of the magnetic field is shown in Figs.~\ref{fig:rho-diag_tau1.0}(a)-\ref{fig:rho-diag_tau1.0}(c), where we set the field angle $\bm B \parallel c$ ($\theta=\phi=0^\circ$) and the band-independent relaxation time $\tau_{\rm h}=\tau_{\rm e}=1.0$ ps.
In the following, we define the field angle $\theta$, $\phi$ as the polar and azimuthal angles, respectively.
The initial increase in diagonal resistivity as a function of the magnetic field $B$ follows a quadratic dependence.
We see the anisotropy $\rho_{yy} < \rho_{xx}$ at $B=0$ T for temperatures $30$ K $< T < 330$ K, contrary to Ref.~\onlinecite{Eo2022c-axis} reporting $\rho_{xx} < \rho_{yy}$, but consistent with a recent study using FIB microstructures that found $\rho_{yy} < \rho_{xx}$ \cite{Zhang2025Dimensionality}.
The anisotropy $\rho_{xx} < \rho_{yy}$ will be obtained by introducing a $\bm k$-dependent relaxation time as shown in Ref.~\onlinecite{Eaton2024Quasi-2D}.
However, this is not within the scope of the present study and will not be further discussed because the angle-resolved magnetoresistance is mainly determined by the warped part of FSs and expected to be insensitive to this $\bm k$-dependence.
The ratio $\rho_{zz}/\rho_{xx} = 8.6$ at $B=0$ T and $T=30$ K is of the same order as the experimental value \cite{Eo2022c-axis}.

The resistivity tensor as a function of the field angle $\theta$ is shown in Figs.~\ref{fig:rho-diag_tau1.0}(d)-\ref{fig:rho-diag_tau1.0}(f) where $B=8.08$ T and $T=90$ K.
The in-plane resistivity $\rho_{xx}$ and $\rho_{yy}$ are monotonically suppressed by increasing angle $\theta$ and are almost independent of $\phi$.
In contrast, we see that the $c$-axis resistivity $\rho_{zz}$ shows a nonmonotonic field angle $\theta$ dependence [Fig.~\ref{fig:rho-diag_tau1.0}(f)]: $\rho_{zz}$ increases monotonically when the magnetic field is rotated from the $c$ axis to the $a$ axis ($\phi=0^\circ$), but weakly oscillates when rotated toward the $b$ axis ($\phi=90^\circ$). However, these behaviors of $\rho_{zz}$ are qualitatively contrary to the experimental results \cite{Kimata2026Electron-Hole}.
%In the experimental results, the $c$-axis resistivity $\rho_{zz}$ monotonically increases with rotating the magnetic field from the $c$ axis to the $b$ axis, while oscillates when rotating the field toward the $a$ axis.

\begin{figure}[tbp]
\includegraphics[width=1.0\linewidth]{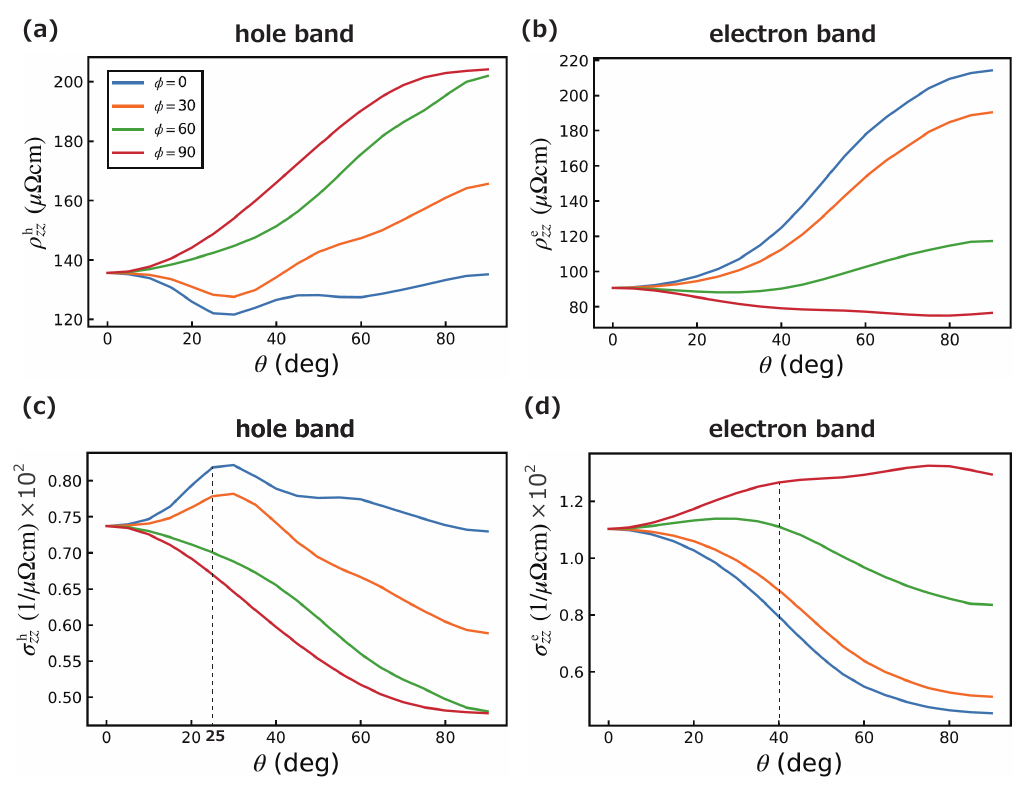}
\centering
\caption{Field angle $\theta$ dependence of band-resolved resistivity [(a), (b)] and conductivity [(c), (d)]. %\YY{per spin}. 
We set the same parameters as Fig.~\ref{fig:rho-diag_tau1.0}(f),  $B=8.08$ T, $T=90$ K, and $\tau_{\rm h}=\tau_{\rm e}=1.0$ ps, and illustrate the results for each field angle $\phi$ by the same color. Dashed lines in (c) and (d) depict the angles $\theta$ adopted in Figs.~\ref{fig:CrossSection}(b) and \ref{fig:CrossSection}(d), respectively.
\label{fig:rho-band_tau1.0}}
\end{figure}

To resolve the discrepancy with the experiment, we discuss the reason behind these behaviors by studying the band-resolved resistivity and conductivity. 
The total conductivity is given by the summation of the hole and electron band components, $\sigma_{ij} = 2(\sigma^{\rm h}_{ij} + \sigma^{\rm e}_{ij})$. Here, factor 2 comes from the spin degree of freedom.
When we consider only the hole band and neglect $\sigma^{\rm e}_{ij}$, we obtain the band-resolved resistivity ${\hat \rho}^{\rm h}(\bm B) = [\hat \sigma^{\rm h}(\bm B)]^{-1}$ %\JIS{$\rho_{zz} = 1/\sigma_{zz}^{\rm h}$} 
in Fig.~\ref{fig:rho-band_tau1.0}(a), which clearly exhibits a trend opposite to Fig.~\ref{fig:rho-diag_tau1.0}(f). %the electron band %\JIS{$\rho_{zz} = 1/\sigma_{zz}^{\rm e}$} 
%[Fig.~\ref{fig:rho-band_tau1.0}(b)].
%At $\phi=0$ $(90)^\circ$,  the hole (electron) band component of $\rho_{zz}$ oscillates with respect to $\theta$. 
At $\phi=0^\circ$, the band-resolved $\rho_{zz}^{\rm h}$ for the hole band oscillates with respect to $\theta$. When $\phi$
is increased, the resistivity increases. In contrast, the band-resolved $\rho_{zz}^{\rm e}$ for the electron band, which is obtained by neglecting $\sigma^{\rm h}_{ij}$, shows similar behaviors to the total resistivity and weakly oscillates at $\phi=90^\circ$. 

%\YYS{Since $\sigma^{\rm h}_{zz}<\sigma^{\rm e}_{zz}$ [see Figs.~\ref{fig:rho-band_tau1.0}(c) and \ref{fig:rho-band_tau1.0}(d)],} 
The above analysis indicates that the field angle dependence of the total resistivity $\rho_{zz}$ is dominated by the electron band component $\sigma^{\rm e}_{zz}$, when the relaxation time is independent of the bands.
In this case, the calculation result of the angle-resolved magnetoresistance [Fig.~\ref{fig:rho-diag_tau1.0}(f)] is qualitatively incompatible with the experiment \cite{Kimata2026Electron-Hole}. 
The analysis of the band-resolved resistivity  naturally leads to an idea. 
When the relaxation time of the hole band is longer than that of the electron band, %we will see $\sigma^{\rm h}_{zz}>\sigma^{\rm e}_{zz}$, 
the transport is dominated by the hole band, and the field angle dependence of magnetoresistance $\rho_{zz}$ can become consistent with the experiment. We examine this idea in the next subsection.

\begin{figure}[tbp]
\includegraphics[width=1.0\linewidth]{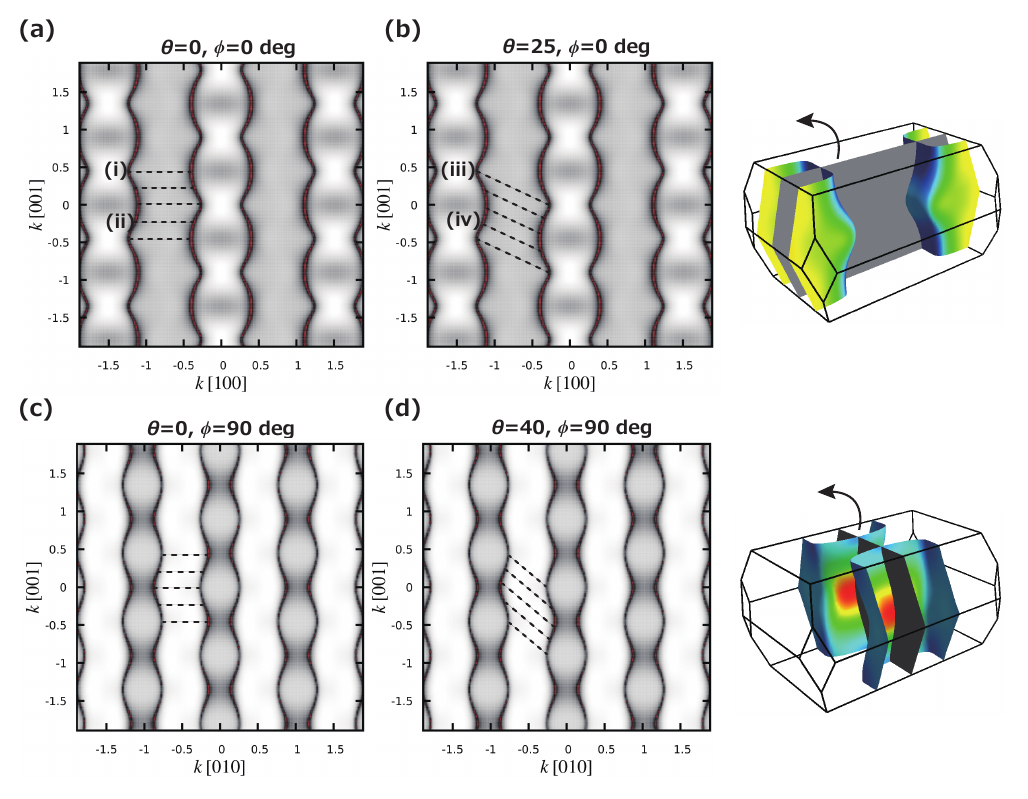}
\centering
\caption{(a), (b) Cross sections of the hole FS in the $k_x$-$k_z$ plane at $k_y=0$. Typical orbits of the hole carriers in the magnetic field oriented along the angle (a) $\theta=\phi=0^\circ$ and (b) $\theta=25^\circ$ and $\phi=0^\circ$ are labeled by (i)-(iv).
(c), (d) Cross sections of the electron FS in the $k_y$-$k_z$ plane at $k_x=0$. Orbits of the electron carriers are illustrated by dashed lines.
\label{fig:CrossSection}}
\end{figure}

To gain an intuitive understanding of the magnetoresitance, here we discuss the origin of the field angle dependence. %in $\sigma^{\rm e}_{zz}$ and $\sigma^{\rm h}_{zz}$. 
The field angle dependence of $\sigma^{\rm e,h}_{zz}$ can be understood by considering the semiclassical trajectories of the carrier and $\overline v_z(\bm k)$.
Figures \ref{fig:CrossSection}(a) and \ref{fig:CrossSection}(b) show the cross sections of the hole FS in the $k_x$-$k_z$ plane at $k_y=0$. 
Typical orbits of hole carriers in the magnetic field oriented along the $c$ axis ($\theta=\phi=0^\circ$) [Fig.~\ref{fig:CrossSection}(a)] and $\theta=25^\circ$ and $\phi=0^\circ$ [Fig.~\ref{fig:CrossSection}(b)] are labeled by (i)-(iv), where the band-resolved conductivity $\sigma^{\rm h}_{zz}$ shows a minimum and maximum, respectively.
In the hole orbit (i), the velocity ${\bm v}(\bm k)$ has only in-plane components, and therefore this orbit does not contribute to the $c$-axis conductivity $\sigma_{zz}^{\rm h}$.
In the other hole orbit (ii), $v_z(\bm k)$ has a different sign on the opposite sides of the warped FS.
This results in the cancellation of $\overline v_z(\bm k)$ and decreases $\sigma_{zz}^{\rm h}$.
When the magnetic field $\bm B$ is rotated toward the $a$ axis, the orbit (i) changes to the orbit (iii) and encounters an extremum of the FS at $\theta=25^\circ$. 
At this angle, the orbit (iii) again gives no contribution to $\sigma_{zz}^{\rm h}$, similar to the orbit (i).
However, in the hole orbit (iv), the trajectory avoids cancellation in contrast to the orbit (ii), resulting in a net increase in $\sigma_{zz}^{\rm h}$.
By further tilting the magnetic field, the trajectories again pass through $\bm k$-points with a sign change of $v_z(\bm k)$ as in the case of $\theta=\phi=0^\circ$ because of the periodicity of the reciprocal lattice, implying a decrease of $\sigma_{zz}^{\rm h}$.
This is a mechanism of the field angular oscillation of the conductivity due to the warping of the FS.
We encounter the same situation for the electron FS when the magnetic field is tilted from the $c$ axis to the $b$ axis [Figs.~\ref{fig:CrossSection}(c) and \ref{fig:CrossSection}(d)]. Because the warping planes are orthogonal between the hole and electron FSs, the field angle dependencies of $\sigma_{zz}^{\rm h}$ and $\sigma_{zz}^{\rm e}$ and the corresponding magnetoresistance show qualitatively different behaviors.

\subsection{Magnetoresistance with band-dependent relaxation time}
\label{sec:result_3}

In the previous section, we assumed band-independent relaxation time and showed that the field angle dependence of magnetoresistance is inconsistent with the experiment for UTe$_2$ \cite{Kimata2026Electron-Hole} because the contribution from the electron band is dominant.
As shown in Figs.~\ref{fig:rho-band_tau1.0}(c) and \ref{fig:rho-band_tau1.0}(d), the hole band component of the conductivity is qualitatively different from the electron band component, and the corresponding resistivity in Fig.~\ref{fig:rho-band_tau1.0}(a) shows the field angle dependence in agreement with the experimental results \cite{Kimata2026Electron-Hole}.
Therefore, we consider the band-dependent relaxation time and assume $\tau_{\rm h}>\tau_{\rm e}$. %and neglect the momentum dependence of $\tau_n(\bm k)$, for simplicity.
A shorter relaxation time of the electron band is consistent with the dHvA experiment, which suggests $\tau_{\rm h}>\tau_{\rm e}$ since the dHvA amplitude of the electron band is much smaller than that of the hole band \cite{Aoki2022First} (see Appendix \ref{ap:relaxation_time}).

The microscopic origin of the band-dependent relaxation time could be attributed to magnetic fluctuations.
In UTe$_2$, antiferromagnetic fluctuation with the wave vector $\bm q \sim (0,\pi,0)$ has been observed in neutron scattering experiments \cite{Duan2020Incommensurate,Knafo2021Low}.
The self-energy correction from this magnetic fluctuation yields a strong quasiparticle damping on the FSs with a nesting vector $\bm q \sim (0,\pi,0)$.
Thus, the relaxation time %$\tau_n(\bm k)$ 
is expected to be short on the FSs around the $k_y = \pm \pi/2$ plane, including the warping part of the electron FS (see Appendix \ref{ap:relaxation_time}). 
For simplicity, we neglect the momentum dependence of the relaxation time $\tau_n(\bm k)$ on each band. This is justified for the following discussions because %the field angle dependence of 
the magnetoresistance is mainly determined by the warped parts of the FSs.

%at the corresponding $\bm k$-points.

\begin{figure}[tbp]
\includegraphics[width=1.0\linewidth]{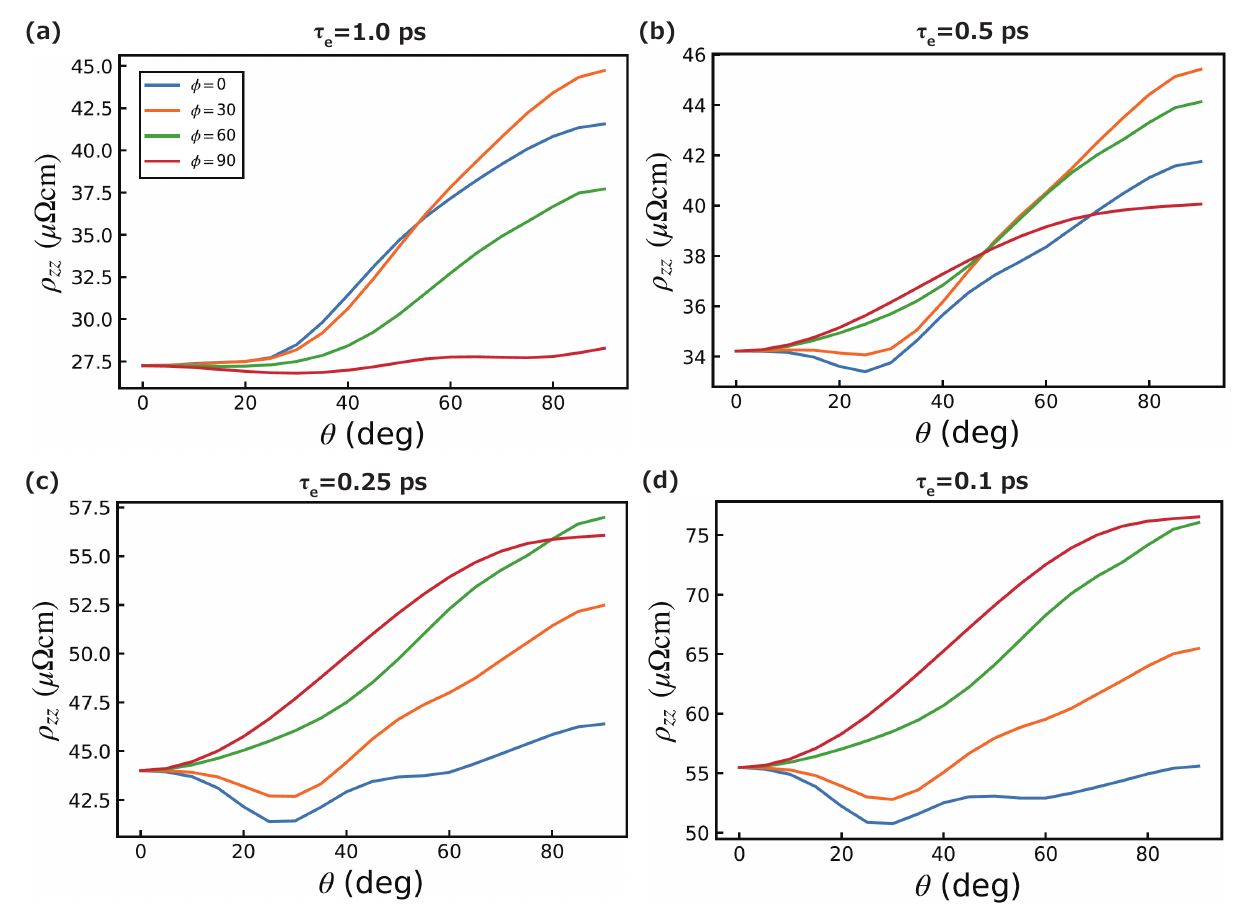}
\centering
\caption{(a)-(d) Field angle $\theta$ dependence of the $c$-axis resistivity with the band-dependent relaxation time. We fix $\tau_{\rm h}=1.0$ ps and vary $\tau_{\rm e}$ from (a) 1.0 ps to (d) 0.1 ps. The other parameters, $B=8.08$ T and $T=90$ K, are the same as Fig.~\ref{fig:rho-diag_tau1.0}(f).
%in each field angle $\phi$.
\label{fig:rho-theta_tau0.1-1}}
\end{figure}

\begin{figure}[htbp]
\includegraphics[width=1.0\linewidth]{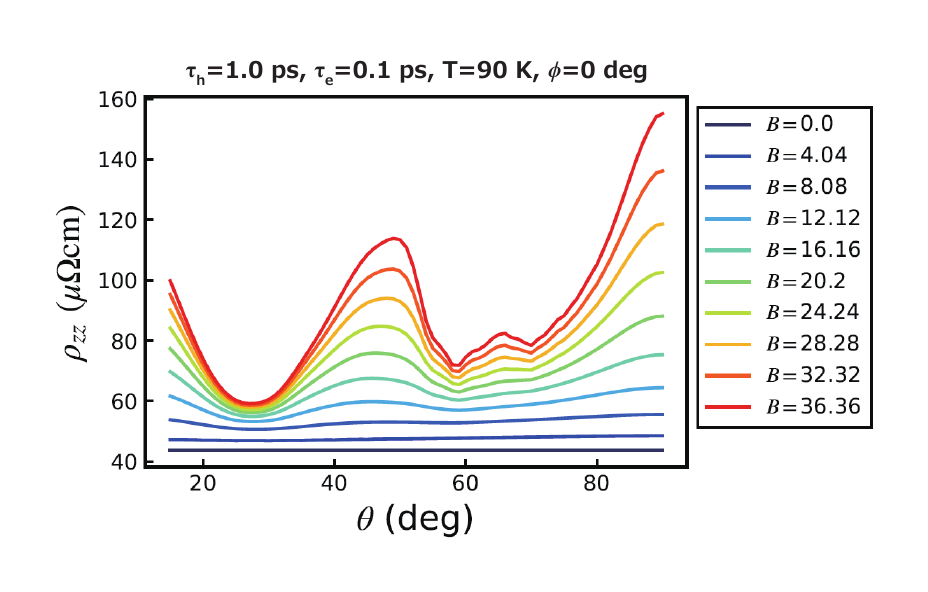}
\centering
\caption{Field angle $\theta$ dependence of the $c$-axis resistivity for $T=90$ K, $\tau_{\rm h}=1.0$ ps, $\tau_{\rm e}=0.1$ ps, and $\phi=0^\circ$. The magnetic field is increased from $B=0$ T to $B=36.36$ T from the bottom to the top.
\label{fig:rho-theta_B0-40}}
\end{figure}

In Fig.~\ref{fig:rho-theta_tau0.1-1}, we show the $\theta$ dependence of the total resistivity $\rho_{zz}$ for several relaxation times $\tau_{\rm e}$ of the electron band. The relaxation time of the hole band is fixed to $\tau_{\rm h}=1.0$ ps, and we assume $B=8.08$ T and $T=90$ K as in Fig.~\ref{fig:rho-band_tau1.0}.
The results for $\tau_{\rm e}=0.25$ and $0.1$ ps appear to be consistent with the experiment \cite{Kimata2026Electron-Hole}. We see in Figs.~\ref{fig:rho-theta_tau0.1-1}(c) and \ref{fig:rho-theta_tau0.1-1}(d) that the resistivity $\rho_{zz}$ is gradually suppressed as the angle $\phi$ decreases.
For $\phi=0^\circ$, the resistivity oscillates as a function of $\theta$, and a first minimum appears at $\theta_{1}=25$--$30^\circ$, in good agreement with the experimental result $\theta_{1}^{\rm expt}=30^\circ$.
We find the second maximum at $\theta_{2}=45$--$50^\circ$, which is also consistent with the experimental observation $\theta_{2}^{\rm expt}=48^\circ$.
Figure~\ref{fig:rho-theta_tau0.1-1} does not show a distinguished peak at higher $\theta$, although peaks at $\theta_{3}^{\rm expt}=65^\circ$ and $\theta_{4}^{\rm expt}=72^\circ$ are experimentally observed.
The discrepancy is partially solved, when we increase the magnetic field $B$. An additional peak appears at $\theta_{3}=65^\circ$ for $B>20$ T (Fig.~\ref{fig:rho-theta_B0-40}), consistent with the experiment.
Thus, the field angle dependence of the resistivity $\rho_{zz}$, including the oscillation in the high angle region, is reproduced and attributed to the properties of the hole FS.
For details of the field angle $\theta$ dependence at low and high magnetic fields, see Appendix \ref{ap:Bdep}. 

Due to the limitation on the number of $\bm k$-points, the magnetoresistance has been calculated at the temperature $T=90$ K higher than the experimental value $T\sim1.5$ K~\cite{Kimata2026Electron-Hole}.
Although calculations down to $T \sim 1.5$ K are computationally demanding, our results from $T = 330$ K to $T = 30$ K indicate that the field-angle dependence is qualitatively robust against lowering the temperature. Since the temperature dependence in the present Boltzmann formulation is introduced through the Fermi distribution function, the main contribution to the conductivity and resistivity comes from electronic states in the vicinity of the Fermi level. Therefore, we expect that the essential features of the calculated angular dependence would remain unchanged even at $T \sim 1.5$ K.

Although we have not explicitly calculated the resistivity and conductivity using the Wannier model at different $U$-values, we expect that the qualitative field-angle dependence associated with the warped hole Fermi surface is robust within the range of $U$ values where the Fermi-surface geometry is consistent with the quantum oscillation results.

\subsection{Off-diagonal transport}
Next, we address the field angle dependence of the transverse (Hall) resistivity. From the symmetry constraint, the Hall resistivity $\rho_{ij}$ vanishes when the magnetic field is aligned with either the $i$ or $j$ axis.
The Hall resistivity in the isotropic 2D model with hole and electron carriers is given by \cite{Zhang2019Magnetoresistance}
\begin{equation}
    \rho_{x y}=-\frac{1}{e} \frac{\left(n_{\rm h} \mu_{\rm h}^2-n_{\rm e} \mu_{\rm e}^2\right) B+\mu_{\rm e}^2 \mu_{\rm h}^2\left(n_{\rm h}-n_{\rm e}\right) B^3}{\left(n_{\rm h} \mu_{\rm h}+n_{\rm e} \mu_{\rm e}\right)^2+\left(n_{\rm e}-n_{\rm h}\right)^2 \mu_{\rm e}^2 \mu_{\rm h}^2 B^2},
    \label{eq:rho_xy}
\end{equation}
for ${\bm B} = B\hat{z}$, indicating that $\rho_{xy}=0$ if $n_{\rm h}=n_{\rm e}$ and $\mu_{\rm h}=\mu_{\rm e}$.
Although UTe$_2$ is a compensated metal, the Hall conductivity becomes finite because of the anisotropic mobility in the orthorhombic structure. 
% \YYS{Actually, if we decompose the Hall resistivity into the hole and electron band components, they do not completely cancel out (Fig.~\ref{fig:rho-band-off-diag}). ???}

\begin{figure}[htbp]
\includegraphics[width=1.0\linewidth]{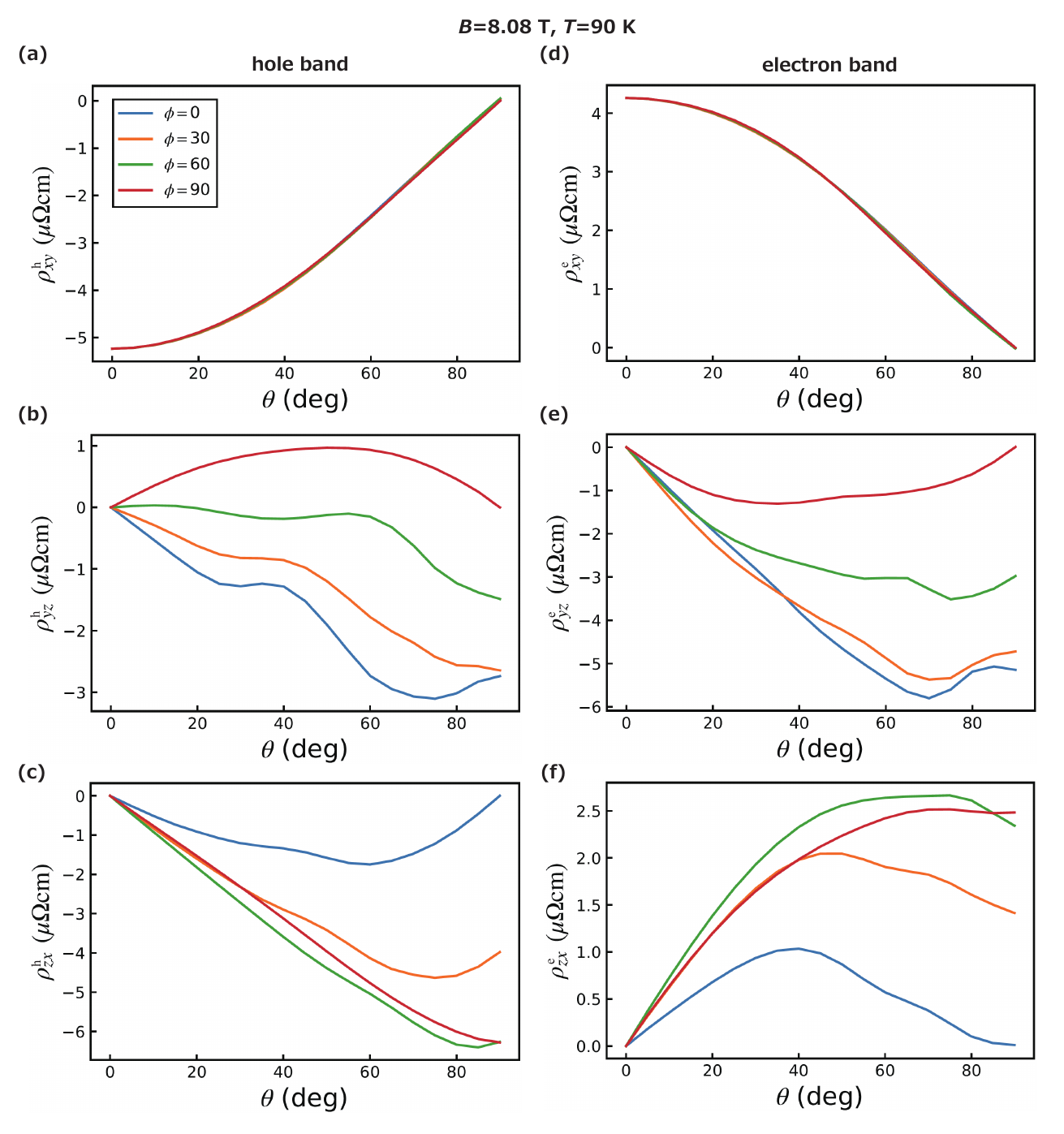}
\centering
\caption{Field angle $\theta$ dependence of the band-resolved Hall resistivity for $B=8.08$ T, $T=90$ K, $\tau_{\rm h}=\tau_{\rm e}=1.0$ ps, and various field angles $\phi$. 
(a)-(c) [(d)-(f)] Only the hole (electron) band is taken into account. 
\label{fig:rho-band-off-diag}}
\end{figure}

\begin{figure}[tbp]
\includegraphics[width=1.0\linewidth]{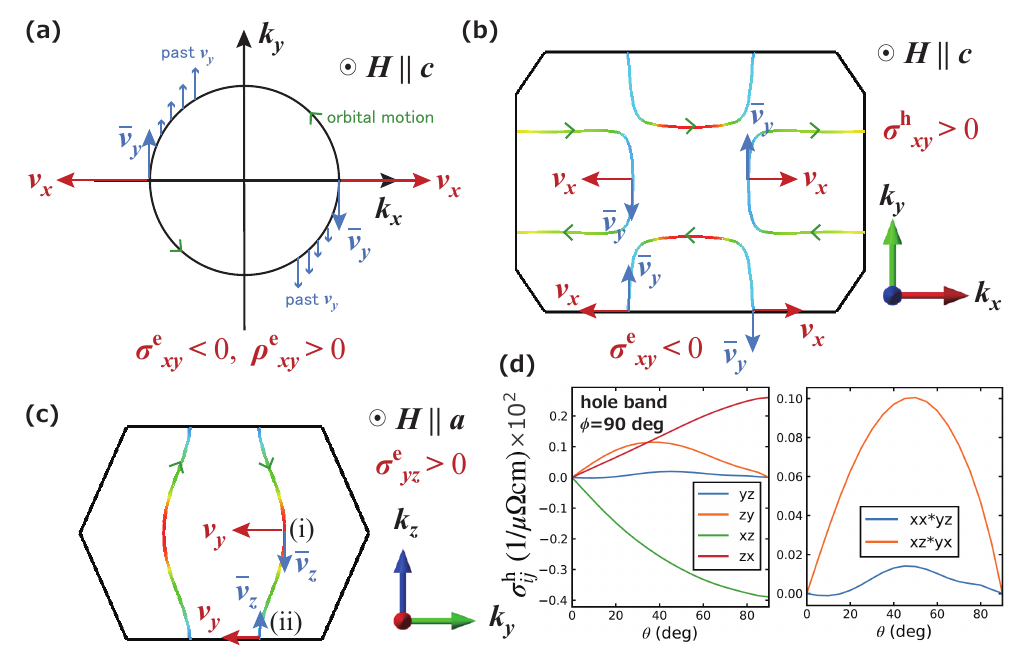}
\centering
\caption{(a) Schematic picture for the Hall conductivity and resistivity in the isotropic 2D free electron model. In the formula derived from the Boltzmann equation, Eq.~\eqref{eq:Boltzmaneq}, $\bar{v}_{y}$ is calculated by integrating the transverse velocity 
$v_{y}$ over the past history ($t<0$) with the weight factor $e^{t/\tau_n}$. 
Because $v_x{\bar v}_{y}<0$ at $\bm k = (\pm k_{\rm F},0)$ that give dominant contribution to $\sigma^{\rm e}_{xy}$, we obtain $\sigma^{\rm e}_{xy}<0$ and $\rho^{\rm e}_{xy}>0$, although a small positive contribution $\sigma^{\rm e}_{xy}(\bm k)>0$ comes from the region $k_xk_y>0$.
(b) Schematic illustration for the Hall conductivity $\sigma^{\rm e,h}_{xy}$ in UTe$_2$ under the magnetic field along the $c$ axis. We can intuitively understand $\sigma^{\rm e}_{xy} < 0$ $\sigma^{\rm h}_{xy} > 0$ as in the isotropic electron and hole models.
(c) Schematic illustration for the Hall conductivity $\sigma^{\rm e}_{yz}$ due to the electron band under the magnetic field along the $a$ axis. 
(d) Left: some components of Hall conductivity $\sigma^{\rm h}_{ij}$ in the hole band. Right: $\sigma^{\rm h}_{xx}\sigma^{\rm h}_{yz}$ and $\sigma^{\rm h}_{xz}\sigma^{\rm h}_{yx}$.
Field angle $\theta$ dependence for $B=8.08$~T, $T=90$~K, and $\phi=90$~deg are plotted.
\label{fig:vk-resolved}}
\end{figure}

% \YYS{The signs of the band-resolved resistivity tensor are consistent with Eq.~(\ref{eq:rho_xy}) and can be intuitively explained by considering $v_i(\bm k)$ and ${\bar v}_j(\bm k)$ in terms of the Boltzmann equation [Figs.~\ref{fig:vk-resolved}(a) and \ref{fig:vk-resolved}(b].} 
We present the Hall resistivity obtained by considering either the hole or electron band in Fig.~\ref{fig:rho-band-off-diag}. The signs of the band-resolved resistivity tensor can be explained by considering $v_i(\bm k)$ and ${\bar v}_j(\bm k)$ in terms of the Boltzmann equation (Fig.~\ref{fig:vk-resolved}). The Hall resistivity $\rho^{\rm e,h}_{xy}$ and $\rho^{\rm e,h}_{zx}$ follow the sign expected from a simple formula Eq.~\eqref{eq:rho_xy}: The electron (hole) band gives positive (negative) Hall resistivity.
However, we obtain a counterintuitive result for $\rho^{\rm h}_{yz} \geq 0$ at $\phi=90^\circ$ [Fig.~\ref{fig:rho-band-off-diag}(b)] and $\rho^{\rm e}_{yz} \leq 0$ [Fig.~\ref{fig:rho-band-off-diag}(e)], which is given by
\begin{align}
\rho^{n}_{yz}=- (\sigma^{n}_{xx}\sigma^{n}_{yz} - \sigma^{n}_{xz}\sigma^{n}_{yx})/\det (\hat{\sigma}^{n}).
\label{eq:rho_yz}
\end{align}
Note that we observe $\rho^{\rm e}_{yz} \leq 0$ in the almost whole parameter space ($B=0$--$40$ T, $T=30$--$300$ K, $\theta=0$--$90^\circ$, $\phi=0$--$90^\circ$). 
This is due to the curvature of the warped electron FS with a large velocity, as illustrated in Fig.~\ref{fig:vk-resolved}(c).
%by considering $v_y(\bm k)$ and ${\bar v}_z(\bm k)$ on the warped plane with the magnetic field along the $a$ axis (Fig.@). 
%The regions (i) and (ii) give positive and negative contribution $\sigma^{\rm e}_{yz}(\bm k)$, respectively.
Because $|{\bm v}(\bm k)|$ is large in the region (i), the $\bm k$-resolved Hall conductivity $\sigma_{yz}^{n}(\bm{k}) \equiv \tau_n v_y^n(\bm{k}) \overline{v}_z^n(\bm{k})\left(-\frac{\partial f}{\partial \varepsilon}\right)_{\varepsilon=\varepsilon_n(\bm{k})}$
in the region (i) is larger than in the region (ii). In Fig.~\ref{fig:vk-resolved}(c), we see that the region (i) gives a positive $\sigma^{\rm e}_{yz}(\bm k)$ as $v_y^{\rm e}(\bm{k}) \overline{v}_z^{\rm e}(\bm{k})>0$.
Since the first term dominates in Eq.~(\ref{eq:rho_yz}), we get $\rho^{\rm e}_{yz} \leq 0$.

There is a different reason for the positive Hall resistivity in the hole band $\rho^{\rm h}_{yz} \geq 0$ at $\phi=90^\circ$.
We find a small Hall conductivity $\sigma^{\rm h}_{yz}$, %a large imbalance between $\sigma^{\rm h}_{ij}$ and $\sigma^{\rm h}_{ji}$, 
as shown in Fig.~\ref{fig:vk-resolved}(d).
This is because the hole FS is elongated along the $k_x$ direction, and the Fermi velocity is small on the warped part of the FS. 
This is in contrast to the electron band, where the warped FS has a large velocity. 
In the hole band, because $\sigma^{\rm h}_{yz}$ is much smaller than $\sigma^{\rm h}_{xz}$, the second term dominates in Eq.~(\ref{eq:rho_yz}) [see right panel in Fig.~\ref{fig:vk-resolved}(d)], and we have $\rho^{\rm h}_{yz} \geq 0$.
% \YYS{while the electron FS is elongated along the $k_y$ direction, with a small Fermi velocity in the opposite direction ($k_x$).
% In other words, the hole FS is more anisotropic than the electron FS regarding the orbital motion.
% Since $v_y(\bm k)$ on the hole FS exists on the flat plane, $\bar v_z(\bm k)\approx0$ gives $\sigma^{\rm h}_{yz}\approx0$. Thus, the second term dominates in Eq.~(\ref{eq:rho_yz}) (Fig.~\ref{fig:vk-resolved}(d)), and we have $\rho^{\rm h}_{yz} \geq 0$.}

\begin{figure}[htbp]
\includegraphics[width=1.0\linewidth]{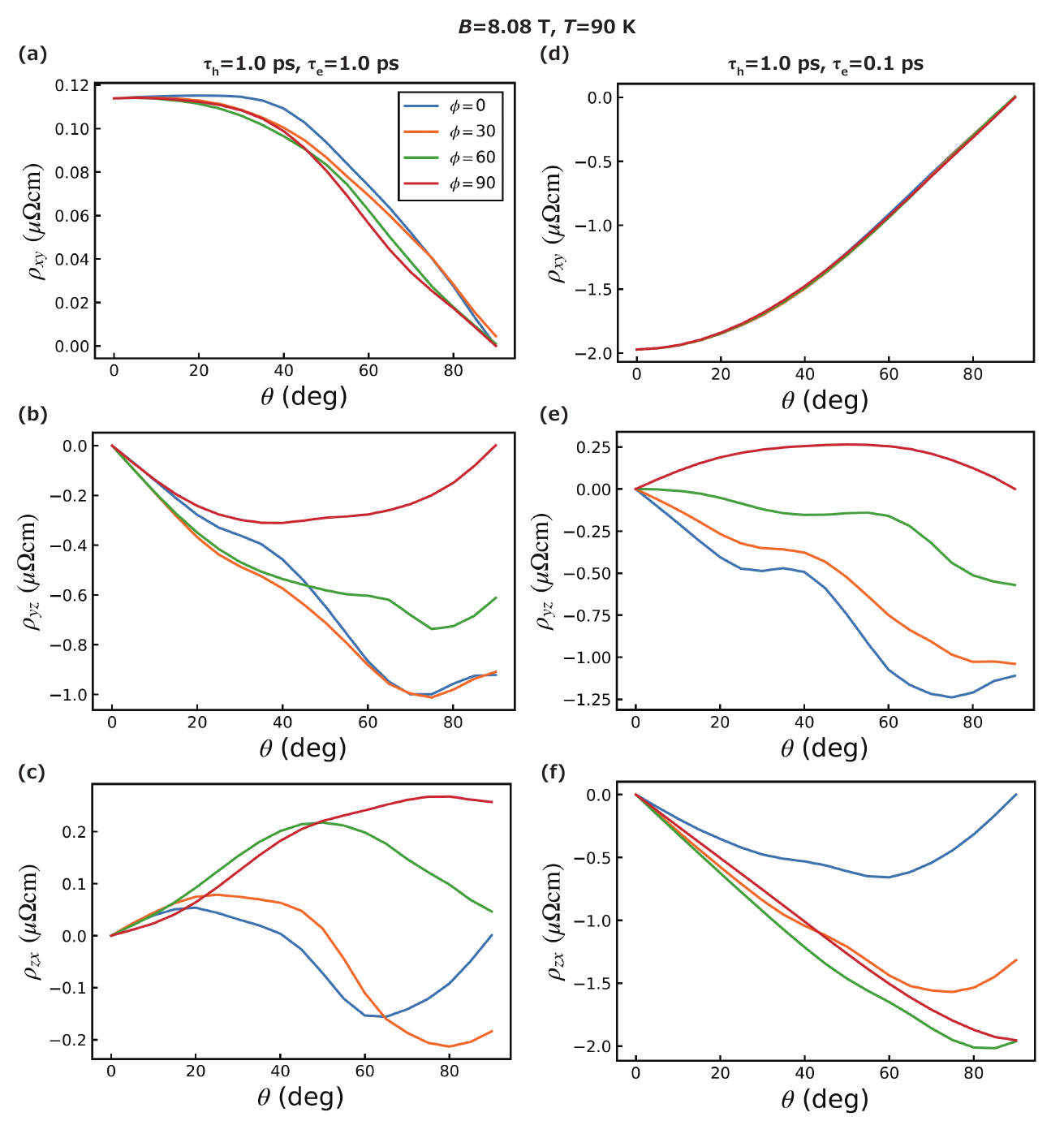}
\centering
\caption{Field angle $\theta$ dependence of the Hall resistivity for $B=8.08$ T and $T=90$ K in each field angle $\phi$. (a)-(c) The results with band-independent relaxation time $\tau_{\rm h}=\tau_{\rm e}=1.0$ ps and (d)-(f) with band-dependent relaxation time $\tau_{\rm h}=1.0$ ps and $\tau_{\rm e}=0.1$ ps.
\label{fig:rho-off-diag}}
\end{figure}

The field angular dependence of the total Hall resistivity is shown in Fig.~\ref{fig:rho-off-diag}, where we set $\tau^{\rm e}=1.0$ or $0.1$~ps. 
For $\tau^{\rm e}=1.0$~ps, we see that the $\theta$ dependence of $\rho_{xy}$ [Fig.~\ref{fig:rho-off-diag}(a)] and $\rho_{yz}$ [Fig.~\ref{fig:rho-off-diag}(b)] is similar to $\rho^{\rm e}_{xy}$ [Fig.~\ref{fig:rho-band-off-diag}(d)] and $\rho^{\rm e}_{yz}$ [Fig.~\ref{fig:rho-band-off-diag}(e)], where transport is determined by the electron band. Similarly to the diagonal resistivity, these off-diagonal components of resistivity are dominated by the electron band when the relaxation time is independent of the band. In contrast, $\rho_{zx}$ shows a complicated $\theta$ dependence due to the contribution of the hole band. 
When we assume a short relaxation time $\tau^{\rm e}=0.1$ of the electron band, as expected, all components of Hall resistivity [Figs.~\ref{fig:rho-off-diag}(d)-(f)] resemble the results for the hole band. 

To our best knowledge, field angle dependence of off-diagonal transport in UTe$_2$ has not been reported.
Thus, we expect that future experimental %observation of sign change in 
studies of off-diagonal transport, compared with our calculation results, will provide further information on the FS and the quasiparticle's relaxation time.

\section{Summary}
We have investigated the magnetic field angular dependence of magnetoresistance in the normal state of the spin-triplet superconductor UTe$_2$.
By applying the semiclassical Boltzmann equation to a 12-band Wannier model, we demonstrated that both the electron and hole bands contribute to the anisotropic magnetoresistance oscillations in the $c$ axis transport with respect to the magnetic field angle. %exhibits angle resolved magnetoresistance oscillations for both hole and electron bands. 
The calculated angular dependence of the magnetoresistance agrees with the experimental results when the relaxation time of the electron band is shorter than that of the hole band, suggesting the importance of the hole band in the electron transport. Electron correlation through magnetic fluctuations is expected to play an essential role in determining the band-dependent relaxation time.
We also find nontrivial behaviors in the Hall resistivity, including a sign change.
The predicted field-angle dependence of the Hall resistivity provides a clear and experimentally testable signature of the contribution from the ordinary intraband transport.
Experimental observation of this phenomenon would serve as a fingerprint of the FS topology.

\begin{acknowledgments}
We appreciate helpful discussions with Motoi Kimata and Toni Helm. Some figures in this paper were created by using VESTA \cite{Momma2011VESTA} and FermiSurfer \cite{Kawamura2019FermiSurfer}. This work was supported by JSPS
KAKENHI (Grant Numbers JP22H01181, JP22H04933,
JP23K17353, JP23K22452, JP24K21530, JP24H00007, JP25H01249, JP25K07223).

\end{acknowledgments}

\appendix
\section{Wannier fitting dependence}
\label{ap:wannierfit}
\begin{figure}[tbp]
\includegraphics[width=1.0\linewidth]{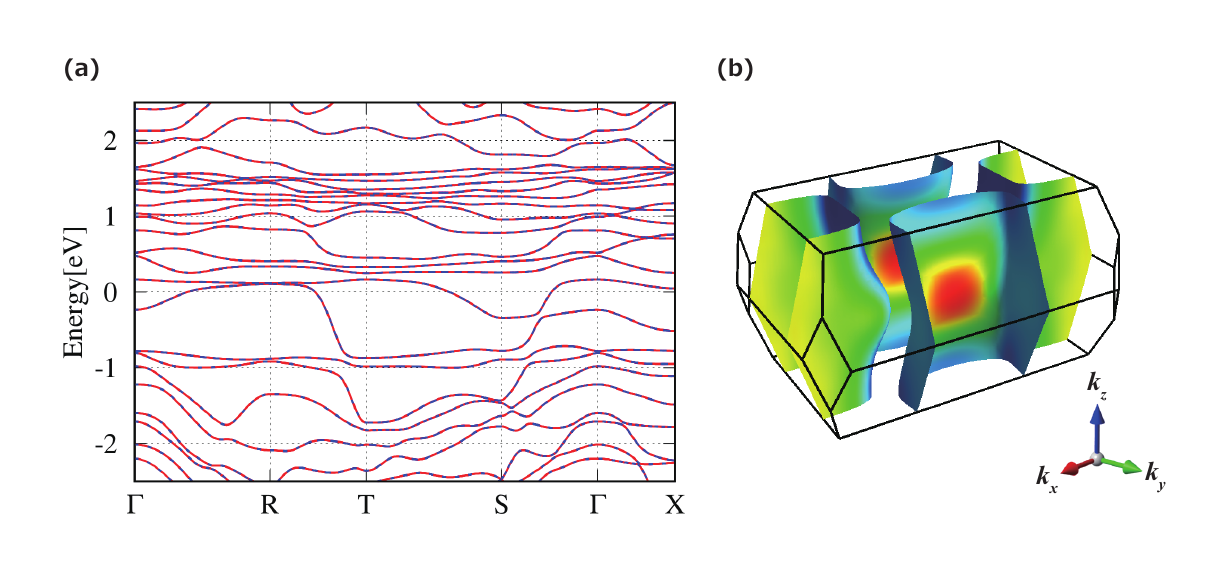}
\centering
\caption{(a) Band structure of the GGA+$U$ calculation for $U=2.0$ eV (solid lines) and the 72-band Wannier model (dashed lines) along the high symmetry lines. (b) Fermi surfaces of the 72-band Wannier model.
\label{fig:model_72band}}
\end{figure}

In Fig.~\ref{fig:model}(b), the agreement between the 12-band Wannier model and the GGA+$U$ ($U=2$ eV) band structure is not fully satisfactory, especially near the $S$-point in the energy range [0, 1] eV. The Wannier fit in UTe$_2$ will be improved by using the 72 projectors for Wannier orbitals (U:5f, U:6d, Te(1):4p, Te(2):4p) as shown in Fig.~\ref{fig:model_72band}. However, a compact model is necessary to reduce the calculation cost of the velocity ${\bm v}^{n}(\bm k (t))$.

\begin{figure}[tbp]
\includegraphics[width=1.0\linewidth]{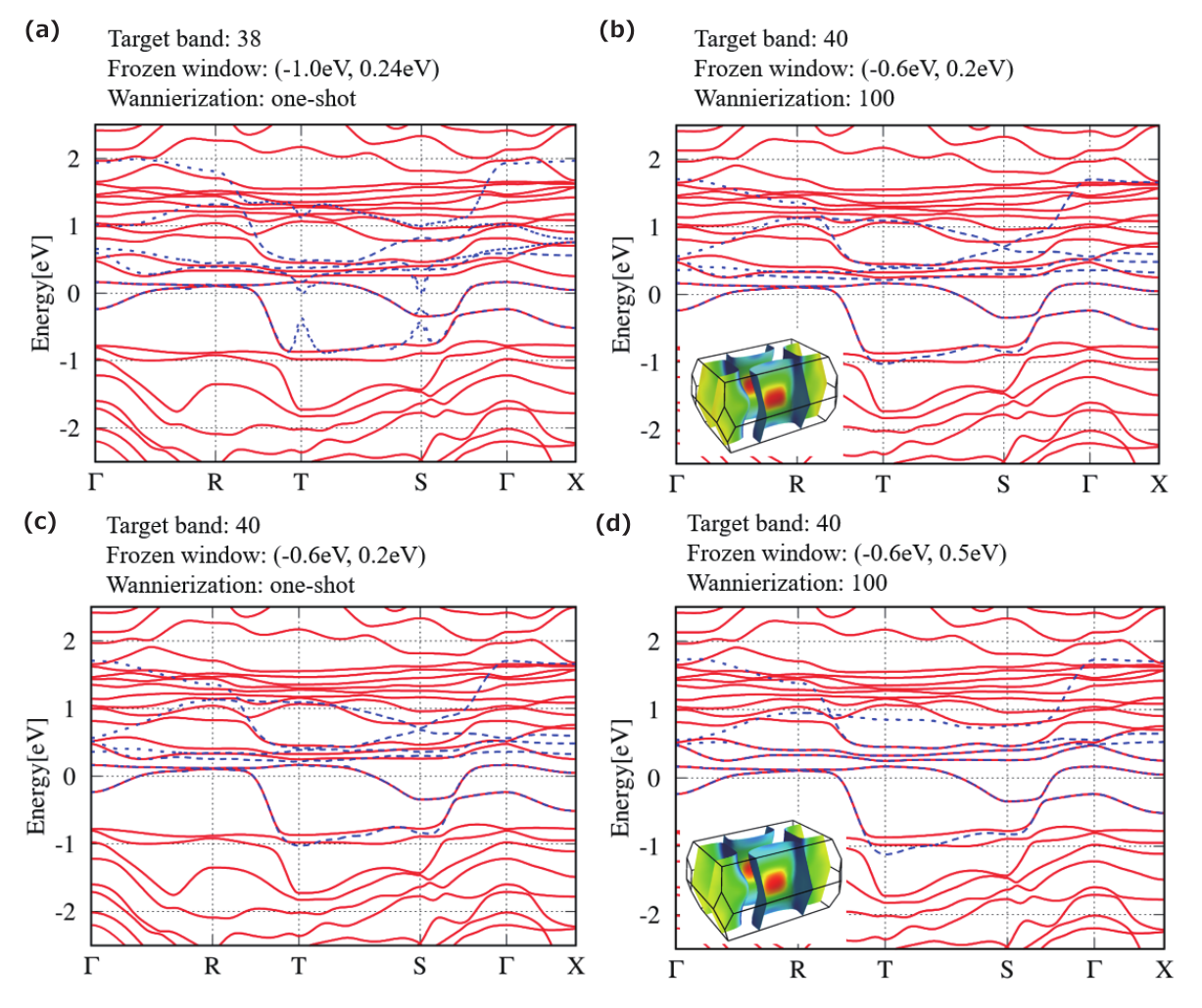}
\centering
\caption{(a)-(d) Band structure of the GGA+$U$ calculation for $U=2.0$ eV (solid lines) and the 12-band Wannier model (dashed lines) using the different set of parameters (frozen energy windows, number of target band, and Wannierization procedure). Inset of (b) and (d) show the Fermi surfaces of corresponding models.
\label{fig:wannierfit}}
\end{figure}

In constructing the 12-band Wannier model, the several parameters (frozen energy windows, number of target band, and Wannierization procedure) are important to improve fitting. Figures~\ref{fig:wannierfit}(a)-(d) show the parameter dependence of the Wannier fit, where the model of Fig.~\ref{fig:wannierfit}(b) is used in the main text. Comparing the Wannier models in Figs.~\ref{fig:wannierfit}(b) and \ref{fig:wannierfit}(d), we find that the choice of the frozen window improves the fitting of the band just above the conduction band near the $S$-point. In the calculations of conductivity and resistivity using the WannierTools package, however, the effects of a band that does not cross the Fermi level are neglected because the derivative of the Fermi distribution function is usually negligible for such bands. Thus, different models are expected to yield essentially the same results when their Fermi surfaces and Fermi velocities are sufficiently similar, as shown in Fig.~\ref{fig:model_72band}(b) and the inset of Figs.~\ref{fig:wannierfit}(b) and \ref{fig:wannierfit}(d). Actually, we have confirmed the quantitative accuracy of present calculations by comparing results of the models in Figs.~\ref{fig:wannierfit}(b) and \ref{fig:wannierfit}(d).

\section{Effects of field amplitude changes}
\label{ap:Bdep}
\begin{figure}[tbp]
\includegraphics[width=1.0\linewidth]{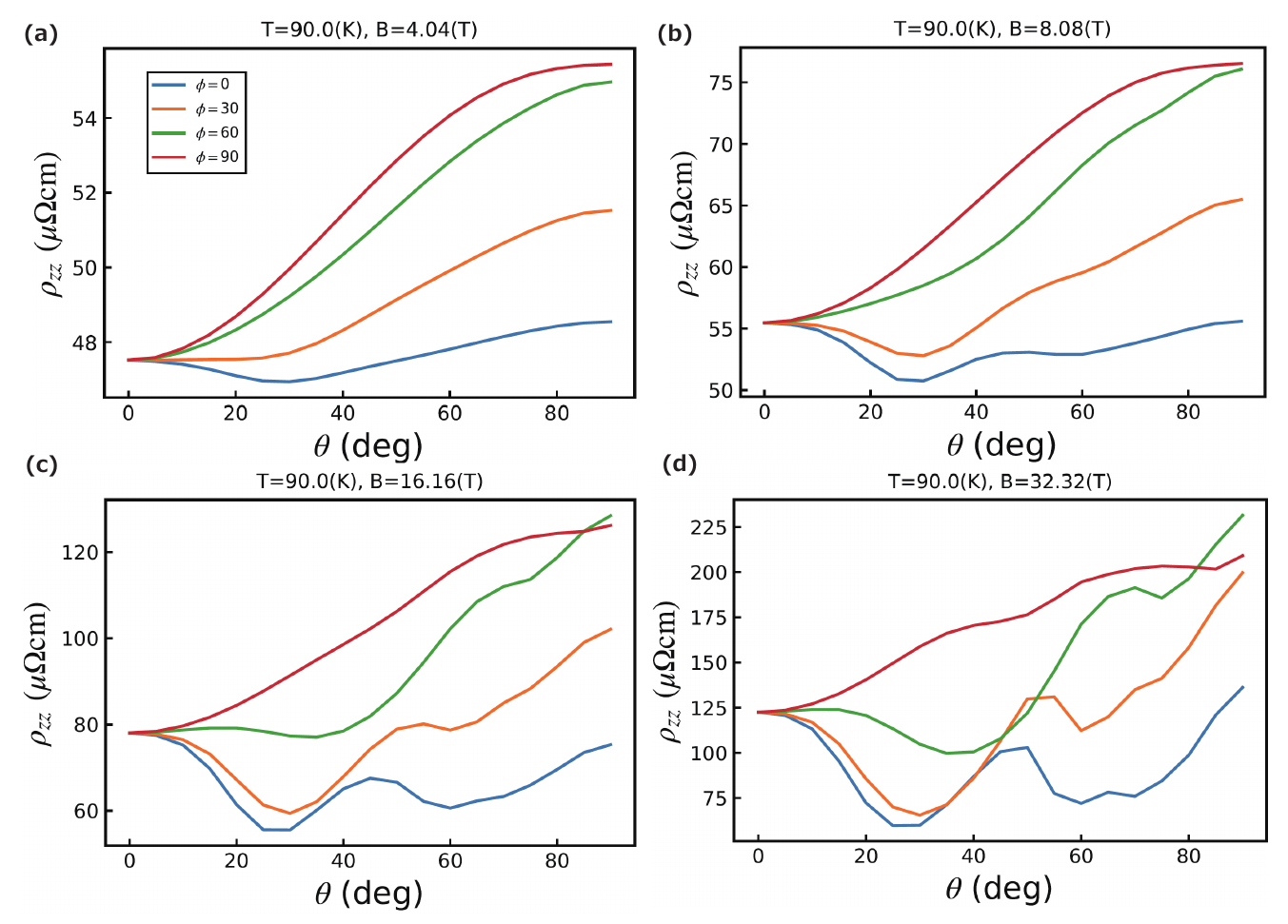}
\centering
\caption{(a)-(d) Field angle $\theta$ dependence of the $c$-axis resistivity with the band-dependent relaxation time. We fix $T=90$ K, $\tau_{\rm h}=1.0$ ps, and $\tau_{\rm e}=0.1$ ps, and vary $B$ from (a) 4.04 T to (d) 32.32 T.
\label{fig:rho-theta_B4-36}}
\end{figure}

To investigate the field angle dependence, we mainly focused on the magnetic field $B=8.08$ T in Sec.~\ref{sec:result}. Figure \ref{fig:rho-theta_B4-36} shows the field angle dependence of resistivity for $B=4.04$, $8.08$, $16.16$, and $32.32$ T. At a low magnetic field $B=4.04$ T, the angle-dependent magnetoresistance oscillation is small [Fig.~\ref{fig:rho-theta_B4-36}(a)]. At a higher magnetic field $B=16.16$ T [Fig.~\ref{fig:rho-theta_B4-36}(c)], the oscillation becomes clearer, but the $\phi=90^\circ$ curve is reduced at $\theta=90^\circ$ and becomes smaller than the $\phi=60^\circ$ curve, which is inconsistent with the experiment (Fig.~2b in Ref.~\onlinecite{Kimata2026Electron-Hole}). In contrast, $B=8.08$ T reproduces the details of the field-angle dependence. In addition, at $\theta=0^\circ$, the magnitude of $\rho_{zz} \sim 55$ $\mu\Omega \rm{cm}$ at $B=8.08$ T is close to the experimental value of $\rho_{zz} \sim 60$ $\mu\Omega \rm{cm}$. Note that at $B=32.32$ T [Fig.~\ref{fig:rho-theta_B4-36}(d)], an additional peak appears at high $\theta$ as discussed in Fig.~\ref{fig:rho-theta_B0-40} of the main text. Overall, although the field-angle dependences at different magnetic fields are qualitatively similar, the results of $B=8.08$ T are shown in the main text, which provide the best agreement with the experimental results.

\section{Relaxation time}
\label{ap:relaxation_time}
\subsection{Estimation of \texorpdfstring{$\tau_{\rm h}$}{tauh} and results for \texorpdfstring{$\tau_{\rm h}>1$}{tauh > 1} ps}
In Sec.~\ref{sec:result}, we set a typical value of relaxation time $\tau_{\rm h}=1$ ps. One possible estimate of relaxation time would be obtained from the quantum oscillation measurements \cite{Aoki2022First}. The Dingle temperature $T_{\rm D} \sim 0.16$ K for hole band may yield $\tau_{\rm h}=7.6$ ps. For electron band, however, such an estimate would be difficult because of the small amplitude of dHvA signal. 

\begin{figure}[tbp]
\includegraphics[width=1.0\linewidth]{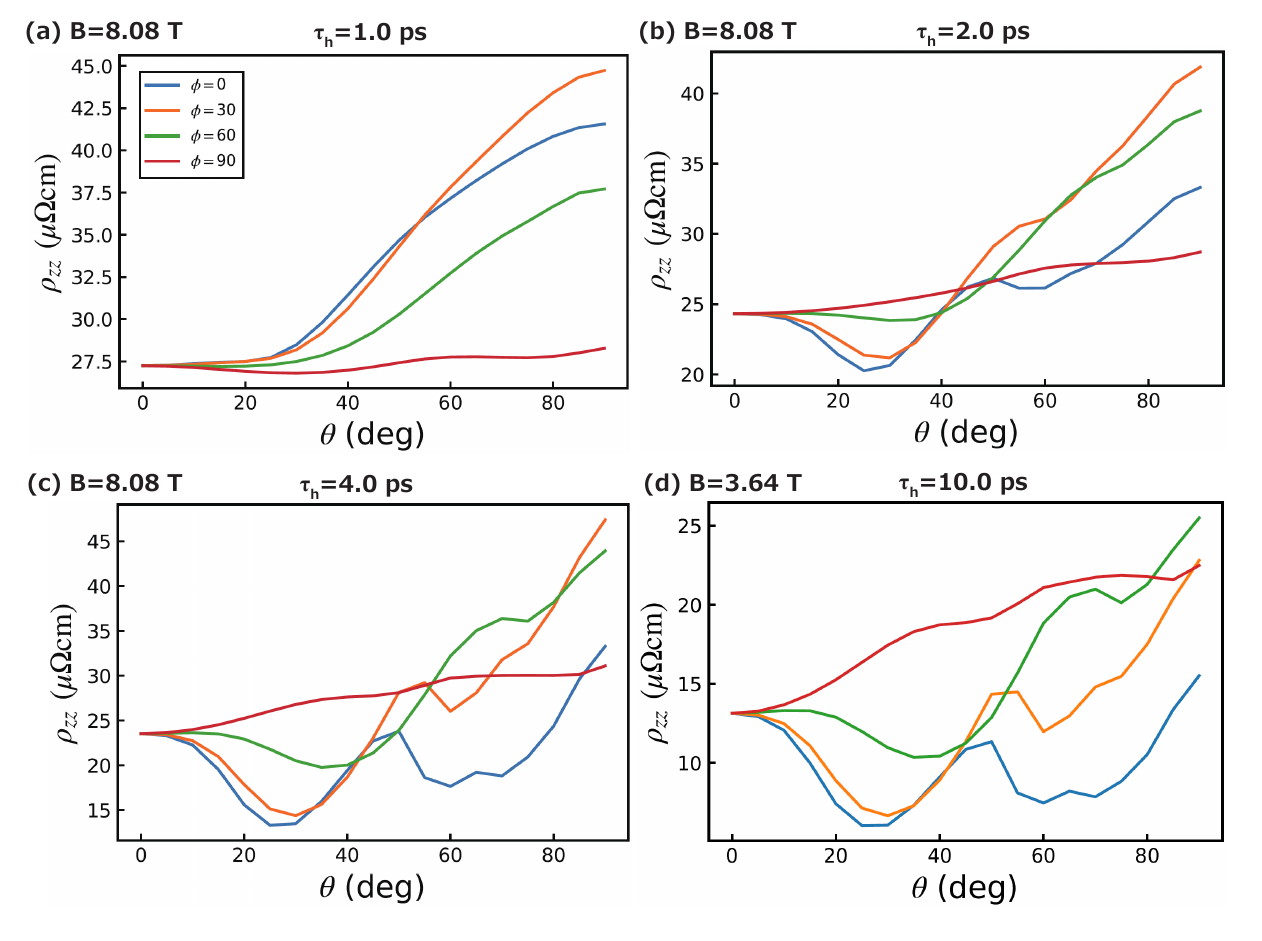}
\centering
\caption{(a)-(d) Field angle $\theta$ dependence of the $c$-axis resistivity with the band-dependent relaxation time. We fix $T=90$ K, $\tau_{\rm e}=1.0$ ps, and (a)-(c) $B=8.08$ T and (d) $B=3.64$ T, and vary $\tau_{\rm h}$ from (a) 1.0 ps to (d) 10.0 ps.
%in each field angle $\phi$.
\label{fig:rho-theta_tau1-10}}
\end{figure}

Figure~\ref{fig:rho-theta_tau1-10} shows the field angle $\theta$ dependence of the $c$-axis resistivity with various relaxation times for hole band longer than $\tau_{\rm h} = 1.0$ ps. We fix $\tau_{\rm e}=1.0$ ps and vary $\tau_{\rm h}$ from $1.0$ ps to $10.0$ ps. We note that the field is set $B=8.08$ T for Fig.~\ref{fig:rho-theta_tau1-10}(a)-(c) and $B=3.64$ T for Fig.~\ref{fig:rho-theta_tau1-10}(d) for computational reasons. From Fig.~\ref{fig:rho-theta_tau1-10}, the robustness of the peak positions and the non-monotonic behaviors are verified, whereas the details of the non-monotonic behaviors and amplitudes are different. Thus, the essentially same results are obtained for increasing $\tau_{\rm h}$ with fixing $\tau_{\rm e}=1.0$ ps and decreasing $\tau_{\rm e}$ with fixing $\tau_{\rm h}=1.0$ ps.

\subsection{An estimation of \texorpdfstring{$\tau_{\rm h}/\tau_{\rm e}$}{tauh/taue}}

A microscopic treatment of relaxation time would be demanding since the calculations of one-particle self-energy and two-particle vertex correction are required. Although we would like to leave such work for future study, we here show that the hierarchy of relaxation times, $\tau_{\rm h} > \tau_{\rm e}$, is obtained from a phenomenological calculation. We estimate the relative transport scattering rate for each band and wave vector $\bm k$ as
\begin{align}
\Gamma_{n}^{\mathrm{tr}}(\boldsymbol{k})&=\frac{1}{N_k} \sum_{m \in \mathrm{FS}} \sum_{\boldsymbol{k}^{\prime}} W_{\boldsymbol{Q}}\left(\boldsymbol{k}, \boldsymbol{k}^{\prime}\right) T_{n \boldsymbol{k}, m \boldsymbol{k}^{\prime}} \nonumber \\
&\times \left(1-\frac{\boldsymbol{v}_{n}(\boldsymbol{k}) \cdot \boldsymbol{v}_{m}(\boldsymbol{k}^{\prime})}{\left|\boldsymbol{v}_{n}(\boldsymbol{k})\right|\left|\boldsymbol{v}_{m}(\boldsymbol{k}^{\prime})\right|}\right) \delta\left(\varepsilon_{m}(\boldsymbol{k}^{\prime})-\mu\right).
\end{align}
Here, we simplify the scattering matrix element as $T_{n \boldsymbol{k}, m \boldsymbol{k}^{\prime}}=\left|\sum_l u_{l n}^*(\boldsymbol{k}) u_{l m}\left(\boldsymbol{k}^{\prime}\right)\right|^2$
where $u_{l m}(\boldsymbol{k})$ is the Wannier eigenvector for U atoms, and
$W_{\bm Q}\left(\boldsymbol{k}, \boldsymbol{k}^{\prime}\right)= \exp \left[-\left|\boldsymbol{k}^{\prime}-\boldsymbol{k}-\boldsymbol{Q}\right|^2 / 2 \sigma_{\bm Q}^2\right]$ represents the enhancement of scattering processes connected by the antiferromagnetic wave vector $\boldsymbol{Q}=(0, \pi, 0)$. The delta function at the Fermi level is approximated by a Gaussian broadening with width $\eta$. $\Gamma_{n}^{\mathrm{tr}}(\boldsymbol{k})$ is proportional to the inverse of relaxation time $1 / \tau_{n}(\boldsymbol{k})$. The relaxation time for hole and electron bands is obtained by taking the weighted average on Fermi surface as $1 / \tau_n=\left\langle 1 / \tau_{n}(\boldsymbol{k})\right\rangle_{\mathrm{FS}}$.

Since the overall magnitude of the scattering matrix element is not determined in this phenomenological treatment, we discuss only the relative band dependence of the relaxation time. This phenomenological procedure gives a larger scattering rate for the electron Fermi surface than for the hole Fermi surface, and the estimated relaxation times satisfy $\tau_{\rm e}/\tau_{\rm h} \sim 0.56$ for $\eta=0.001$ and $\sigma_{\bm Q}=0.1$. Therefore, the present estimate supports the physical plausibility of the band-dependent relaxation time adopted in this work, although a fully microscopic calculation of the self-energy and vertex correction remains an important future problem.

\end{document}